# Automated Extraction of Pharmacokinetic Parameters from Structured XML Scientific Articles: Enhancing Data Accessibility at Scale

Remya Ampadi Ramachandran[1,2,3], Lisa A. Tell[4], Sidharth Rai[1], Nuwan Millagaha Gedara[1], Hossein Sholehrasa[1,2,5], Jim E. Riviere[1,2], Majid Jaberi-Douraki[1,2,3, *]

[1]1DATA Consortium, www.1DATA.life, Kansas State University Olathe, Olathe, KS, USA.

[2]Food Animal Residue Avoidance and Databank Program (FARAD), Kansas State University Olathe, Olathe, KS, USA.

[3]Department of Mathematics, Kansas State University, Manhattan, KS, United States

[4]FARAD, Department of Medicine and Epidemiology, School of Veterinary Medicine, University of California-Davis, Davis, CA

[5]Department of Computer Science, Kansas State University, Manhattan, KS, United States

* Corresponding Author: jaberi@k-state.edu

# Abstract

In the field of pharmacology and drug development, there is a notable absence of centralized, comprehensive, and up-to-date repositories of pharmacokinetic (PK) data. This poses a significant challenge for research and development (R&D) as it can be a time-consuming and challenging task to collect all the required quantitative PK parameters from diverse scientific publications. This quantitative PK information is predominantly organized in tabular format, mostly available as XML, HTML, or PDF files within various online repositories and scientific publications, including supplementary materials. This makes tables one of the crucial components and information elements of scientific or regulatory documents as they are commonly utilized to present quantitative information. Extracting data from tables is typically a labor-intensive process, and alternative automated machine learning models may struggle to accurately detect and extract the relevant data due to the complex nature and diverse layouts of tabular data. The difficulty of information extraction and reading order detection is largely dependent on the structural complexity of the tables. Efforts to understand tables should prioritize capturing the content of table cells in a manner that aligns with how a human reader naturally comprehends the information. The Food Animal Residue Avoidance data bank (FARAD) team has been manually extracting tabular data and other information from literature and regulatory agencies for over 40 years. However, there is now an urgent need to automate this process due to the large volume of publications released daily. The accuracy of this task has become increasingly challenging, as manual extraction is tedious and prone to errors, especially given the staffing shortages we are currently facing. This necessitates the development of AI algorithms for table detection and extraction that are able to precisely handle cells organized according to the table structure, as indicated by column and/or row header information. The objective of this paper is to address the knowledge gap by introducing a system that automates the extraction and formatting of data from tables found in published documents, while also taking reading order detection into account. The proposed system encompasses a comprehensive approach to table data extraction that systemically addresses a wide range of both simple and complex layouts. Our results show that the extraction system can effectively and efficiently collect PK data from tables in the XML format of scientific literature.

# Introduction

The primary objective of pharmacology and drug development is to discover and optimize therapeutic compounds that exhibit favorable pharmacokinetic (PK) and pharmacodynamic (PD) properties. Both in human and veterinary medicine, there is a compelling emphasis on comprehending the PK properties of chemical substances and pharmaceutical drugs. The PK properties of a drug play a critical role in the drug discovery process, as a poorly designed study with an incorrect dosing threshold can significantly jeopardize the overall safety profile as well as the efficacy of a new drug and may lead to its termination during the clinical development phase [1,2]. The druggability of a compound largely hinges on its metabolism and PK properties. These aspects pose significant challenges in the field of pharmaceutical research and development (R&D) [3]. Additionally, there is a notable lack of centralized, comprehensive, and up-to-date repositories for PK data within the drug development pipeline. This absence makes it time-consuming and difficult to gather the necessary quantitative PK parameters from various scientific publications. For over 40 years, the Food Animal Residue Avoidance Databank (FARAD) team has manually extracted tabular data and other information from literature and regulatory agencies. However, with the increasing volume of daily publications, there is an urgent need to automate this process. Manual extraction has become increasingly challenging, as it is tedious and prone to errors, especially given the current staffing shortages we are facing.

Improving PK predictions of novel chemical entities or drugs can be achieved by leveraging knowledge and understandings gathered from previous studies conducted on other compounds. However, the process of searching for, curating, and standardizing the relevant PK information from scientific literature still presents a constraint in the development of robust algorithms aimed at enhancing the preclinical and clinical phases of drug development. Collecting and compiling all the essential quantitative PK parameters from scientific publications for R&D purposes is a challenging task. Furthermore, the quantitative PK data in scientific publications is commonly presented within tables, including supplementary data, which can often exhibit complexity and can be challenging to interpret accurately. This complexity further complicates the task of detecting and converting the information into a scientifically meaningful order.

These tables can be quite complex, to the extent that even human readers may struggle to fully comprehend and interpret their content. Therefore, the cognitive ability of the readers, their understanding of the material, and their knowledge acquisition may all be affected [4,5]. However, relying exclusively on manual data curation can impose limitations on our capacity to keep pace with the continuously expanding body of veterinary and biomedical literature. In the field of veterinary and biomedical research, the PubMed database witnesses an influx of more than 1 million papers every year, averaging approximately 2 papers every minute. As a result, the database has now amassed a repository of over 35 million published records including citations and abstracts, thereby complicating and delaying the process of discovering pertinent studies [6–10]. In addition, the extraction of table data from scientific articles across various databases results in the creation of extensive file collections, making the data extraction process even more labor intensiveness, challenging, and exhausting [11].

When it comes to literature mining, augmenting text mining workflows with table data extraction holds the potential to significantly improve performance by granting access to more structured or systematic datasets [12–16]. However, this field necessitates the development of innovative techniques such as dynamic and statistical text and data extraction. Natural language processing (NLP) and text mining offer various tools and methods for retrieving relevant information [17–19]. In recent years, large language models (LLMs) have demonstrated the ability to efficiently and accurately extract data from published records and transform it into a structured format. These models have also shown the potential to address challenges faced in data extraction using NLP techniques. Various studies have confirmed that by leveraging NLP and LLMs, a robust repository of prior published data from scientific articles can be established [20–23]. However, the full capabilities of these LLMs in comprehending data from tables with diverse layouts, heterogeneous data, named entity recognition and classification, relationship identification and classification, and metric system identification and classification, still need further exploration [24,25]. Table size also plays a critical role. Larger tables make it difficult for LLMs to navigate back to previously viewed information, and the presence of multiple headers can further worsen usability challenges [24,25]. Extracting information from tables remains a challenge due to the lack of a systematic approach to handle diverse structures of tables found in literature. This can lead to reduced performance efficiency and difficulty in collecting relevant table information tailored to specific requirements. As part of the layout challenges, these workflows may also struggle to extract information from complex tables with spanning cells across rows, columns, or both. It is important to develop an algorithm capable of effectively managing merged rows and/or columns in data table, ensuring proper distribution, detecting the reading order, and accurately extracting information. However, creating such an algorithm is challenging due to the diverse array of table formats that exist. Automated systems encounter challenges when it comes to efficiently interpret this information [26–28]. Reading order detection is a crucial component of these techniques as it ensures that the data is interpreted in the same way as human readers comprehend it [29]. When it comes to extracting information from multi-dimensional tables, a step-by-step approach is necessary. This begins with identifying the table itself, followed by analyzing its function, structure, and meaning. Finally, the data can be extracted from each individual cell of the table using syntactic processing. This method has been shown to have a higher level of performance efficiency [17].

**Methodology**

**System overview**

An overview of the table data extraction method presented in this study is shown in *Figure 1*. The process begins with a repository of articles in XML format gathered through our web crawler project [11]. Each article is then analyzed to extract tables, converted to desired layouts during the data collection phase. The data from these tables is curated into sentences, and each value is mapped to its corresponding PK parameters of interest. The results are then consolidated into a presentable format such as data frames or in CSV file. To elaborate further, main sources of data is the 1DATA article repository of XML versions of scientific literature published in the PK aspects of Anatomical Therapeutic Chemical (ATC) code QD, QJ, QH, QP [11]. Here, the ATC code QD corresponds to dermatologicals, QJ corresponds to anti-infectives for systemic use, QH corresponds to steroid anabolic growth promoters from systemic hormonal preparations excluding sex hormones and insulin, QP corresponds to antiparasitic products, insecticides, and repellents, and other drugs like ionophores [11,30]. This repository has files from different article providers such as Scopus, Springer, CrossRef through the corresponding API calls. The XML versions of scientific articles are mainly considered for this study. In the absence of XML articles, corresponding HTML files can be used when available.

The overall method consists of a combination of different processes including table detection, reading order detection, PK parameter identification, and PK data extraction. Once the table structure is identified by the proposed algorithm, then data recognition and extraction where the PK parameters were annotated based on the pharmacokinetics ontology is performed [31]. PK parameters, including drugs, AUCs, half-life, $C_{max}$, $T_{max}$, units, volume of distributions, clearances, matrices, and routes, are the key factors considered here. However, different names, including abbreviations and expansions, have been used to refer to parameters such as, CL = clearance, t1/2 = half-life, AUC = area under the concentration curve, and AUCR = AUC ratio as shown in *Table 1* —This has caused them to undergo various iterations to conform to a standard pattern. In this way, quantitative measures such as dose, duration, and PK parameters were identified and extracted as a sequence of *parameter name, unit, value*. The XML table extraction methods described in this context are primarily concerned with four distinct table formats that are specifically related to PK analysis. To facilitate automated handling of these table layouts, the initial phase of the table extraction algorithm involves searching for table indices.

**Table 1:** Representative PK parameters and their alternatives used in the table data extraction module.

| PK Parameters | Some Alternatives (case-insensitive) |
|---|---|
| Clearance | clearance\|cl\|cleared\|cl/f\|cl-f |
| Area under the plasma concentration time curve | area under the curve\|AUC\| area under the concentration curve |
| Time to peak drug concentration | Tmax\|$T_{max}$\|T.?max)\|(time.*Cmax)\|(Time (of\|to) Maximum concentration |
| Ratios of maximal plasma concentration | Cmax\|$C_{max}$\|C max |
| Terminal half-life | Half-life\|half life\|t1/2)\|t 1/2\|T½ |

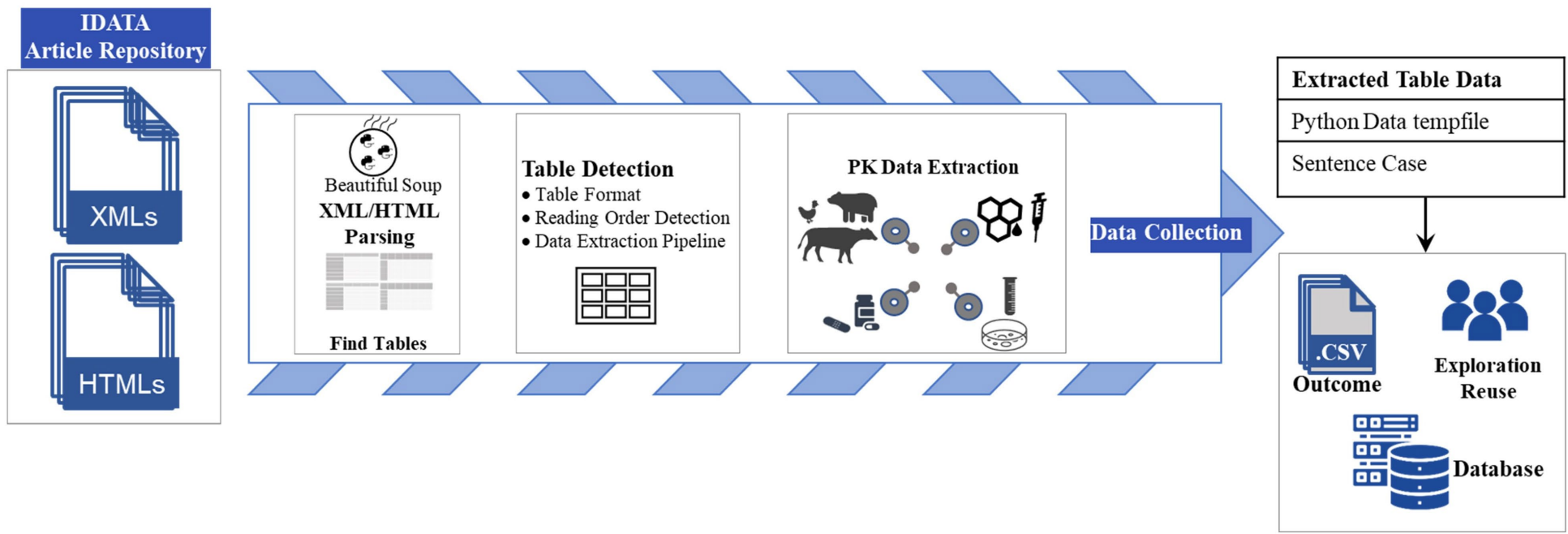

Figure 1: Overall workflow of table data extraction methodology from scientific articles of XML and HTML file types.

**Table Data Extraction**

The first step in extracting PK parameters from tables involved identifying and extracting the tabular forms from various XML file types of scientific articles, obtained from the 1DATA article repository. The success of this process hinged on the selection of appropriate tags to locate and retrieve the tables, which could differ depending on the article provider used (e.g., Scopus, Springer, CrossRef APIs). For the XML files from Scopus database, the proposed algorithm primarily focused on identifying tables designated with the tag element *'ce:table'*. To construct a tabular framework, information about the number of columns and rows in each table was extracted using table elements such as '*tgroup*' and '*row*', respectively. By separating the table headers and body using XML tags like '*thead*' and '*tbody*', the PK parameters in each table could be more easily understood. A row-wise or column-wise sweep was conducted to extract the entry tag elements in each cell and embed them into the output table framework. This same process was applied to files from other article providers, with corresponding XML tag options for table detection.

The module for extracting table data automatically has a dedicated process for HTML files. While the same method can be used for HTML files, it is treated as a separate module to avoid any complications when dealing with multiple file types and tags at once.

**Diversity of the Tables**

The tables present in scientific articles frequently show a complex logical structure of merged-column, merged-row, with spanning cells (as shown in *Table 2*), and each group may have its own header. The table structure detection module is designed to recognize the internal structure of a table, which is a key step in converting tables comprehensible by machines. In a situation like this, such as when dealing with XML files, it is important to not only determine the number of columns and rows using tgroup and row elements, but also accurately identify where merged-columns and merged-rows occur. This is achieved by selecting the appropriate tag elements, such as *namest* and *nameend* for merged-columns, and *morerows* for merged-rows, within the table cell entry elements. By utilizing these tags, the number of merged-columns and merged-rows can be calculated, resulting in the creation of a comprehensive table structure (*Table 2a–b*). Likewise, based on the number of merged cells identified in column-wise and row-wise traversal, proposed algorithm splits spanning cells to replicate the desired cell content. Here, the spanning cells refers to combining multiple cells into single larger cell, enabling a single cell to stretch across multiple columns or rows [32,33]. In order to prevent any inconsistencies in string matching, XML tags for special character styles like italics, bold, superscript, and subscript that are used in the table fields are not considered. To avoid any errors in PK parameter recognition, we addressed the instances of empty cells by inserting 'NaN' in place of them.

**Table 2**: a) An example of a complex table with spanning cells in column and row data fields, and b) complete table structure generated by using the proposed algorithm.

a)

```
+-----------+---+---+---+
| A         | B | C | D |
|           +---+---+---+
|           |     E     |
+---+---+---+---+---+---+
|   | c | c | c | c | c |
+ I +---+---+---+---+---+
|   | c | c | c | c | c |
+---+---+---+---+---+---+
| J | c | c | c |   |   |
+---+---+---+---+---+---+
```

b)

```
+---+---+---+---+---+---+
| A | A | A | B | C | D |
+---+---+---+---+---+---+
| A | A | A | E | E | E |
+---+---+---+---+---+---+
| I | c | c | c | c | c |
+---+---+---+---+---+---+
| I | c | c | c | c | c |
+---+---+---+---+---+---+
| J | c | c | c |NaN|NaN|
+---+---+---+---+---+---+
```

**Automated Reading Order Detection of Table Formats**

As mentioned in the previous section, scientific articles that are accessible online present numerous table structures. By processing these tables independently, it is possible to extract information effectively. Moreover, a reliable data extraction algorithm should automatically identify these table formats without any loss of data. This module is implemented to determine specific table format and redirect it to the desired table data extraction algorithm. The selection of data parsing libraries for this study included evaluating third-party packages such as *lxml* [34], *Beautiful Soup* [35], and *PyQuery* [36]. Given the future requirement to integrate both HTML and XML scientific articles for table data curation, the *Beautiful Soup* package was chosen for further use. Thus, the initial part of table pipeline is implemented using *Beautiful Soup*, a Python package for performing XML or HTML parsing [35]. With the parsed information, this module examines the table indices (both rows and columns) to identify the table format. Outcome will be flagged depending on the table format and triggers the desired table pipeline for content retrieval, as explained in the subsequent section.

For example, upon receiving an XML file for table extraction, the program will analyze the content of the file and identify the different table formats that are present. As representative examples,

a)

**Table 1**
Pharmacokinetic parameters of MEL after IV administration at 0.5 mg/kg in lactating goats (n = 6).

| Parameter | Units | IV Mean SD |
|---|---|---|
| AUC | h·ng/mL | 26499 ± 4233 |
| K10 | 1/h | 0.12 ± 0.03 |
| K12 | 1/h | 0.64 ± 0.38 |
| K21 | 1/h | 1.13 ± 0.71 |
| K10_HL | h | 6.07 ± 1.18 |
| Alpha | 1/h | 1.82 ± 1.09 |
| Beta | 1/h | 0.07 ± 0.02 |
| Alpha_HL | h | 0.53 ± 0.35 |
| Beta_HL | h | 9.96 ± 2.51 |
| A | ng/mL | 1223 ± 153.71 |
| B | ng/mL | 1840 ± 357.69 |
| AUMC | h·h·ng/mL | 374373 ± 120223 |
| MRT | h | 13.88 ± 3.36 |
| CL | mL/h/kg | 19.38 ± 3.86 |
| Vss | mL/kg | 262.37 ± 50.74 |
| V1 | mL/kg | 165.76 ± 23.06 |
| V2 | mL/kg | 96.61 ± 31.07 |

Area under the curve (AUC), elimination rate from compartment 1 (K10), rate of movement from compartment 1–2 (K12), the rate of movement from compartment 2–1 (K21), half-life of the elimination phase (K10_HL), rate constant associated with distribution ($\alpha$), rate constant associated with elimination ($\beta$), distribution half-life (Alpha_HL), elimination half-life (Beta_HL), intercept for the distribution phase (A), intercept for the elimination phase (B), area under the first moment curve (AUMC), mean resident time (MRT); total clearance (CL), volume of distribution at the steady state (Vss), volume of compartment 1 (V1), and volume of compartment 2 (V2).

b)

**Table 6**
PK Parameters Predicted by PBPK Models of Losartan and Carboxylosartan After IV Administration

| Substrate and Dose | PBPK Projections | | Observed[b] |
|---|---|---|---|
| | GastroPlus[a] | Simcyp[a] | |
| Losartan dosed at 20 mg | | | |
| Losartan | | | |
| $V_{ss}$ (L/kg) | 0.17 | 0.24 | 0.38 ± 0.18 |
| CL (mL/min/kg) | 7.2 | 7.0 ± 1.3 | 9.1 ± 1.5 |
| $t_{1/2}$ (h) | 1.95 | 0.63 ± 0.11 | 1.8 ± 0.6 |
| $AUC_{(0-inf)}$ ng.h/mL | 797 ± 97.3 | 744 ± 144 | 539 ± 95 |
| Carboxylosartan | | | |
| $C_{max}$ (ng/mL) | 75.0 | 145 ± 64.7 | 76 ± 30 |
| $AUC_{(0-inf)}$ (ng.h/mL) | 710 | 1075 ± 535 | 744 ± 272 |
| $AUC_m/AUC_p$ | 0.9 | 1.4 | 1.4 |
| Losartan dosed at 30 mg | | | |
| Losartan | | | |
| $V_{ss}$ (L/kg) | 0.17 | 0.24 | 0.62 ± 0.27 |
| CL (mL/min/kg) | 7.2 | 7.0 ± 1.3 | 8.0 ± 2.7 |
| $t_{1/2}$ (h) | 1.95 | 0.63 ± 0.11 | 3.2 ± 0.8 |
| $AUC_{(0-inf)}$ ng.h/mL | 1197 ± 146 | 1095 ± 212 | 851 ± 253 |
| Carboxylosartan | | | |
| $C_{max}$ (ng/mL) | 108 | 214 ± 95.5 | 149 ± 79 |
| $AUC_{(0-inf)}$ (ng.h/mL) | 1083 | 1588 ± 790 | 1332 ± 490 |
| $AUC_m/AUC_p$ | 0.9 | 1.5 | 1.6 |
| Carboxylosartan dosed at 20 mg | | | |
| $V_{ss}$ (L/kg) | 0.16 | 0.21 | 0.14 ± 0.02 |
| CL (mL/min/kg) | 0.36 | 0.72 ± 0.17 | 0.67 ± 0.08 |
| $t_{1/2}$ (h) | 5.0 | 3.73 ± 0.79 | 6.3 ± 1.4 |
| $AUC_{(0-inf)}$ ng.h/mL | 7192 ± 1215 | 7355 ± 1648 | 5969 ± 709 |

[a] Simulated mean results ± SD.
[b] Observed mean data ± SD from study.[9]

c)

**Table 1**
Selected pharmacokinetic parameters describing the disposition of topical eprinomectin in plasma and milk after oral administration (0.5 and 1 mg/kg BW) to lactating goats. Values are the mean ± standard deviation [$C_{max}$: observed peak plasma concentration; $t_{max}$: time to reach $C_{max}$; $t_{1/2\,elim}$: half life of elimination; MRT: mean residence time; AUC: area under the plasma or milk concentration vs time curve].

| Dose | 0.5 mg/kg BW (n=5) | | 1 mg/kg BW (n=5) | |
|---|---|---|---|---|
| Parameters | Plasma | Milk | Plasma | Milk |
| $C_{max}$ (ng/ml) | 15.48 ± 6.64[a] | 5.34 ± 2.23[A] | 38.10 ± 8.57[b] | 11.47 ± 2.23[B] |
| $T_{max}$ (d) | 0.5 | 0.5 | 0.5 | 0.5 |
| $T_{1/2\,elim}$ (d) | 1.36 ± 0.38[a] | 1.33 ± 0.37[A] | 0.95 ± 0.17[a] | 0.99 ± 0.08[A] |
| MRT (d) | 1.02 ± 0.12[a] | 0.97 ± 0.19[A] | 1.12 ± 0.15[a] | 1.11 ± 0.14[A] |
| AUC (ng d/ml) | 17.62 ± 9.68[a] | 6.56 ± 4.00[A] | 45.32 ± 13.90[b] | 13.88 ± 1.77[B] |
| AUC milk to plasma ratio | | 0.36 ± 0.05[a] | | 0.33 ± 0.08[a] |
| Dose fraction recovered in milk | | 0.43 ± 0.19[a] | | 0.42 ± 0.06[a] |

Means in the same row with different small-letter (a, b) or capital-letter (A, B) superscripts are significantly different ($P < 0.05$).

d)

**Table 3**
PK of Carboxylosartan in the Dog Following IV Administration

| ID | Half-life (h) | $V_{ss}$ (L/kg) | $AUC_{0-24hr}$ (ng.h/mL) | $AUC_{0-\infty}$ (ng.h/mL) | $CL_p$ (mL/min/kg) | $f_e$ | $CL_{r,Dog}$ (mL/min/kg) |
|---|---|---|---|---|---|---|---|
| Dog1 | 13 | 1.5 | 1020 | 1040 | 16 | 0.28 | 4.5 |
| Dog2 | 7.4 | 1.1 | 790 | 790 | 21 | 0.25 | 5.3 |
| Mean | 10 | 1.3 | 900 | 920 | 19 | 0.27 | 5.0 |

**Figure 1**: Common table formats. These tables were obtained from the following XML documents: a) 10.1016/j.smallrumres.2018.01.001, b) 10.1016/j.xphs.2017.03.032, c) 10.1016/j.vetpar.2015.02.013, d) 10.1016/j.xphs.2017.03.032.

articles including '10.1016/j.smallrumres.2018.01.001' [37], '10.1016/j.xphs.2017.03.032' [38], and '10.1016/j.vetpar.2015.02.013' [39], were considered where we have the common table structures (*Figure 2*).

In this module, the program will check table indices, row and column headers, and the presence of spanning cells to confirm the table format, then control will be handed over to the table retrieval section based on the table format. For example, for the table formats presented in *Figure 2a–d)*, different cases as detailed in the subsequent sections will be invoked. In this way, with *Beautiful Soup* Python package and layout determination algorithm, table structure is decided to invoke the corresponding table data extraction module. An intermediate phase in this sequence involves handling of spanning cells to replicate the cell value based on the number of columns or rows present in the original table. This specific section holds significant importance as it can reproduce a table without merged-columns or merged-rows. By eliminating any cells that span across multiple rows or columns, it creates a more comprehensive and organized table.

**PK Content Retrieval for Different Table Formats**

Once the table structure has been established and a finalized data frame format has been created, the subsequent step is to extract necessary information from the fields or cells. Here, information is extracted by row-wise or column-wise traversal based on the table layout and generates a *parameter-value-unit* combination for the selected PK parameters. The outcome of this algorithm is a reading order sentence format that helps in easy information extraction to the final .csv file format (*Figure 3*).

*Case 1: Common table format*

This module functions by extracting table content by row-wise traversal while appending header row to each cell value. In this way, a sentence format is generated for each row of the table facilitating easier information extraction. An example of tables considered under this category is shown in *Figure 4a* and program extracted table is given in *Figure 4b*. This category is considered as one of the commonly used table formats where we have the PK parameter names listed as first column and information of column content or index is available in the first row of the table. Other important aspects of this module include identifying and extracting any drug name in the table title and appending at the end of extracted row data.

*Case 2: Common table format - Transpose*

Second category of tables also has very straight forward structure similar to Case 1, but its column appears to be transposed when compared to Case 1. In this case, to retrieve table information as a sentence, a column-wise reading option is activated and append the drug details at the end of it.

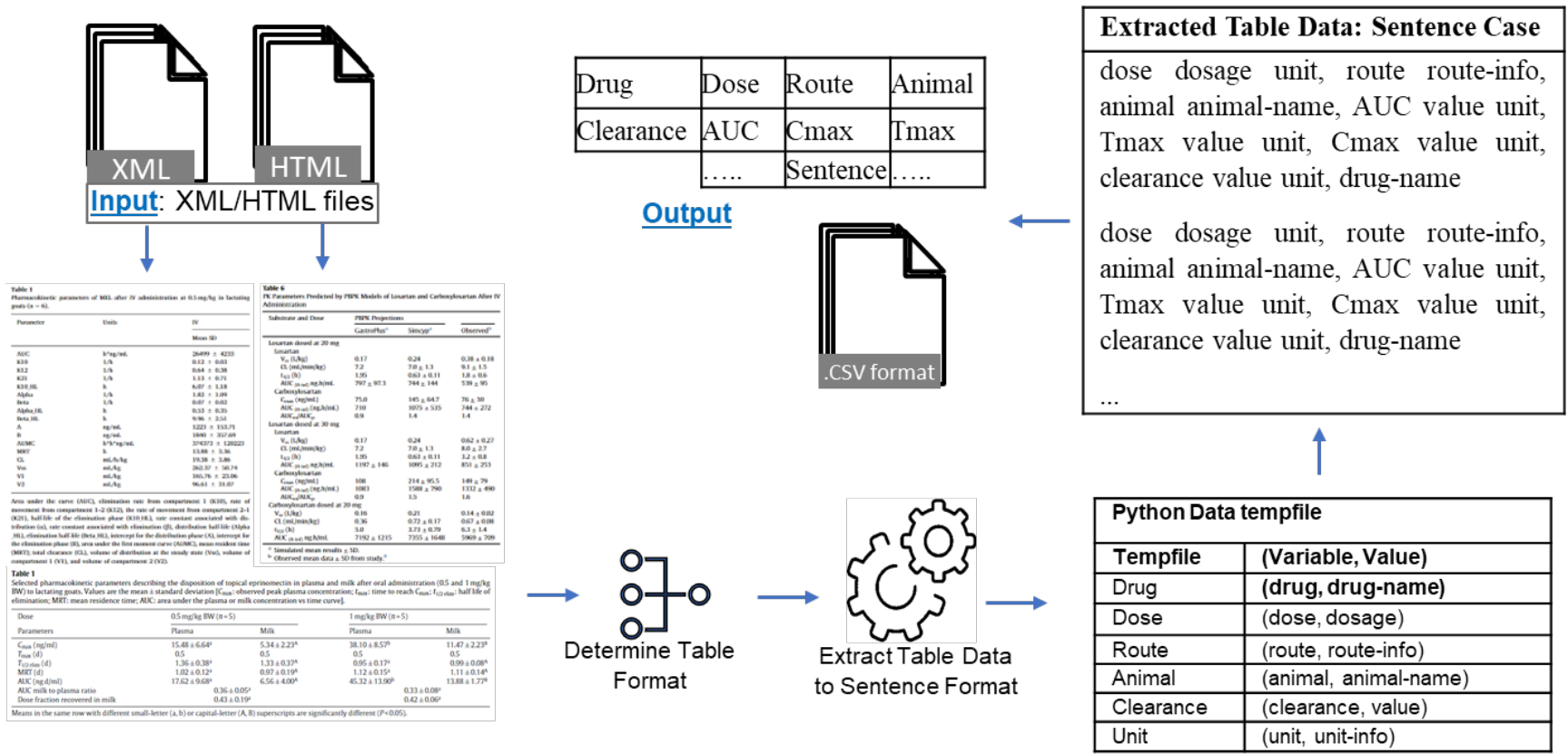

**Figure 2:** Flow chart representing the systematic workflow of table extraction algorithms from scraping the tables to extracting results in a .csv file format for research purposes.

If these conditions are not met, the table will instead be parsed column by column based on the table's specific format. When dealing with these two scenarios, the first step is to locate the table's indices and title. From there, the table will be examined row by row, extracting the information as a string with spaces in between.

*Case 3: Tables with merged column/spanning cells as header row*

Tables having merged columns in the header row is a bit complicated as it can lead to information loss if not managed carefully (*Figure 5a*). This scenario is addressed by using a product function that replicates the merged header information to match with the original data columns (*Figure 5b*). It will generate a table having common structure (*Figure 5c*) which can be managed by row-wise traversal while appending the dosage and matrix information at the end of the sentence, followed by drug name from the table title.

*Case 4: Tables with merged rows/spanning cells in the index column*

Similar to case 3, algorithm to handle merged rows in the tables also uses a product function to replicate the missing information in the index column. Once the table structure is normalized, an algorithm that deals with column-wise table extraction is invoked to generate the sentence pattern. Here also, the dose information, matrices from the index column, and drug information from the table title will be appended at the end of each sentence.

For this, after identifying the table indices and title, merged cells are replicated based on the number columns available in the subsequent rows (*Figure 5a-b*). Then a product function is enabled to parse the table, detect, and extract parameters in a sequence and generates the *Outcome*.

*Case 5: An advanced automated approach*

In addition to managing cases on their own, a more sophisticated option has been created to extract table data from numerous XML files all at once and with complete automation. It gathers all the cases from case 1 to case 4 under a single scheme with the possibility of automatic table format determination and extraction of table data. To accomplish this task, the initial step is to iterate over table headers for drug, animal, matrix data, and PK parameters. For any single occurrence of these names, information corresponding to it is retrieved and stored in variables for subsequent usage. Once the table header is determined, the program will iterate over each cell, decide on the variable category present, and then retrieve data and unit information, until the end of a row is reached. The same process is continued for all the rows present in the table and generates a meaningful sentence with a combination of PK parameter, value, unit, matrix, dose, route, and drug information. In another scenario, where multiple drugs, animals, or matrices are found in the table header, the initial data storage is restricted. Such information is handled when we iterate over each cell to retrieve table data to generate a meaningful sentence from each row. Another possibility with this method is the extraction of table data based on user needs. For example, if the PK data mining research is focused on clearance data, the PK parameters that are extracted from the tables include drug, dose, animal, route, and clearance values. Altogether, this method gives more flexibility to handle table data extraction irrespective of its outline.

## Results

Thousands of full-text scientific articles in XML file formats have been considered for table extraction in this study [11]. As discussed in the methodology section, XML parsing is performed using *Beautiful Soup* where it exclusively looks for available tables. Then for each table its layout is identified, and the desired table data extraction module is invoked. Results for the same articles *'10.1016/j.smallrumres.2018.01.001'; '10.1016/j.vetpar.2015.02.013'* discussed in the methodology section are detailed here.

As a result, one of the tables from '*10.1016/j.smallrumres.2018.01.001*' is identified as common table format with the possibility of row-wise parsing. Parsed results for the selected table are shown in *Figure 4*, having PK Parameter, unit, value, followed by header line and drug name present in the table title, for each row. An example of table identified using *Beautiful Soup*'s *findAll* method for the XML file version is shown in *Figure 4a*. While extracting the table blank content has been replaced with 'NaN' for easy handling. Content retrieved here is in the same structured format, however, for data curation text and data mining studies unstructured data will be more helpful (*Figure 4b*). Considering this fact, further data extraction is planned with focus given on extracting the PK parameters and its interventions matching the identifiers that are set in the defined procedure. It will generate an outcome in an unstructured manner, following a row-wise curation criterion (*Figure 4c*). For tables that have a transposed format, the curation path applied will be column-wise generating unstructured sentence case outcome. The result of table information extracted as a sentence which has drug name MEL from the table title is appended at the end of each row data. [Note: ‖ indicates separation between table rows].

For the article '10.1016/j.vetpar.2015.02.013', table formats available need to be considered under Case 3 scenario. Structured data outcome obtained for one of the tables present in this article is given below in *Figure 5*. Here the parsed string has PK parameter, unit, value, followed by its dosage value and unit, matrix value, from the header row, and drug name from the table title. *Figure 5a* shows the retrieved table using *Beautiful Soup*'s *findAll* method for the article. Initial scenario replicates the spanning cells in the content-data row based on the number of columns, while the spanning cells in the header row is handled with the indexing option. Resulting tables that have structured data format is shown in *Figure 5b–c*. In order to convert it to an unstructured form, a product function is employed as discussed in the Case 3 method section. Data extraction here also focuses on PK parameters and its interventions matching the identifiers that are set in the defined procedure and generate an unstructured outcome (*Figure 5d*). Tables (Case 4) with a transposed structure of Case 3 are handled by considering the product from PK parameters in the table header row and matrices in the index column.

Every table within the XML file is processed in the same manner. It is transformed into a string format according to the specific table format and subsequently sent to the content retrieval module. Here all the information from the table is recorded in a .csv file with details such as PK parameters, units, values, followed by the entire string which is row-wise or column-wise parsed table content. An example of csv result file for the Scopus XML articles is given in *Supplementary Material S1*.

a)

**Table 1**
Pharmacokinetic parameters of MEL after IV administration at 0.5 mg/kg in lactating goats (n = 6).

| Parameter | Units | IV |
|---|---|---|
| | | Mean SD |
| AUC | h*ng/mL | 26499 ± 4233 |
| K10 | 1/h | 0.12 ± 0.03 |
| K12 | 1/h | 0.64 ± 0.38 |
| K21 | 1/h | 1.13 ± 0.71 |
| K10_HL | h | 6.07 ± 1.18 |
| Alpha | 1/h | 1.82 ± 1.09 |
| Beta | 1/h | 0.07 ± 0.02 |
| Alpha_HL | h | 0.53 ± 0.35 |
| Beta_HL | h | 9.96 ± 2.51 |
| A | ng/mL | 1223 ± 153.71 |
| B | ng/mL | 1840 ± 357.69 |
| AUMC | h*h*ng/mL | 374373 ± 120223 |
| MRT | h | 13.88 ± 3.36 |
| CL | mL/h/kg | 19.38 ± 3.86 |
| Vss | mL/kg | 262.37 ± 50.74 |
| V1 | mL/kg | 165.76 ± 23.06 |
| V2 | mL/kg | 96.61 ± 31.07 |

Area under the curve (AUC), elimination rate from compartment 1 (K10), rate of movement from compartment 1–2 (K12), the rate of movement from compartment 2–1 (K21), half-life of the elimination phase (K10_HL), rate constant associated with distribution (α), rate constant associated with elimination (β), distribution half-life (Alpha _HL), elimination half-life (Beta_HL), intercept for the distribution phase (A), intercept for the elimination phase (B), area under the first moment curve (AUMC), mean resident time (MRT); total clearance (CL), volume of distribution at the steady state (Vss), volume of compartment 1 (V1), and volume of compartment 2 (V2).

b)

| | Parameter | Units | IV |
|---|---|---|---|
| | NaN | NaN | Mean SD |
| 2 | AUC | h*ng/mL | 26499 ± 4233 |
| 3 | K10 | 1/h | 0.12 ± 0.03 |
| 4 | K12 | 1/h | 0.64 ± 0.38 |
| 5 | K21 | 1/h | 1.13 ± 0.71 |
| 6 | K10_HL | h | 6.07 ± 1.18 |
| 7 | Alpha | 1/h | 1.82 ± 1.09 |
| 8 | Beta | 1/h | 0.07 ± 0.02 |
| 9 | Alpha_HL | h | 0.53 ± 0.35 |
| 10 | Beta_HL | h | 9.96 ± 2.51 |
| 11 | A | ng/mL | 1223 ± 153.71 |
| 12 | B | ng/mL | 1840 ± 357.69 |
| 13 | AUMC | h*h*ng/mL | 374373 ± 120223 |
| 14 | MRT | h | 13.88 ± 3.36 |
| 15 | CL | mL/h/kg | 19.38 ± 3.86 |
| 16 | Vss | mL/kg | 262.37 ± 50.74 |
| 17 | V1 | mL/kg | 165.76 ± 23.06 |
| 18 | V2 | mL/kg | 96.61 ± 31.07 |

c)

AUC  h*ng/mL  26499 ± 4233 Parameter  NaN  Units  NaN  IV  Mean SD. MEL ‖ K10  1/h  0.12 ± 0.03 Parameter  NaN  Units  NaN  IV  Mean SD. MEL ‖ K12  1/h  0.64 ± 0.38 ‖ K21  1/h  1.13 ± 0.71 Parameter  NaN  Units  NaN  IV  Mean SD. MEL ‖ K10_HL  h  6.07 ± 1.18 Parameter  NaN  Units  NaN  IV  Mean SD. MEL ‖ Alpha  1/h  1.82 ± 1.09 Parameter  NaN  Units  NaN  IV  Mean SD. MEL ‖ Beta  1/h  0.07 ± 0.02  Parameter  NaN  Units  NaN  IV  Mean SD. MEL ‖  Alpha_HL  h  0.53 ± 0.35 Parameter  NaN  Units  NaN  IV  Mean SD. MEL ‖ Beta_HL  h  9.96 ± 2.51 Parameter  NaN  Units  NaN  IV  Mean SD. MEL ‖ A  ng/mL  1223 ± 153.71 Parameter  NaN  Units  NaN  IV  Mean SD. MEL ‖ B  ng/mL  1840 ± 357.69 Parameter  NaN  Units  NaN  IV  Mean SD. MEL ‖  AUMC  h*h*ng/mL  374373 ± 120223 Parameter  NaN  Units  NaN  IV  Mean SD. MEL ‖ MRT  h  13.88 ± 3.36 Parameter  NaN  Units  NaN  IV  Mean SD. MEL ‖ CL  mL/h/kg  19.38 ± 3.86 Parameter  NaN  Units  NaN  IV  Mean SD. MEL ‖ Vss  mL/kg  262.37 ± 50.74 Parameter  NaN  Units  NaN  IV  Mean SD. MEL ‖ V1  mL/kg  165.76 ± 23.06 Parameter  NaN  Units  NaN  IV  Mean SD. MEL ‖  V2  mL/kg  96.61 ± 31.07  Parameter  NaN  Units  NaN  IV  Mean SD. MEL

**Figure 3:** Case 1: a) Table extracted from the following XML document: https://doi.org/10.1016/j.smallrumres.2018.01.001, b) program extracted table, c) Result of table information extracted as a sentence which has drug name MEL from the table title, appended at the end of each row data. [Note: ‖ indicates separation between table rows] Sentence format outcome with parameter, value, units.

a)

Table 6
PK Parameters Predicted by PBPK Models of Losartan and Carboxylosartan After IV Administration

| Substrate and Dose | PBPK Projections | | Observed[b] |
|---|---|---|---|
| | GastroPlus[a] | Simcyp[a] | |
| Losartan dosed at 20 mg | | | |
| Losartan | | | |
| $V_{ss}$ (L/kg) | 0.17 | 0.24 | 0.38 ± 0.18 |
| CL (mL/min/kg) | 7.2 | 7.0 ± 1.3 | 9.1 ± 1.5 |
| $t_{1/2}$ (h) | 1.95 | 0.63 ± 0.11 | 1.8 ± 0.6 |
| AUC$_{(0-inf)}$ ng.h/mL | 797 ± 97.3 | 744 ± 144 | 539 ± 95 |
| Carboxylosartan | | | |
| $C_{max}$ (ng/mL) | 75.0 | 145 ± 64.7 | 76 ± 30 |
| AUC$_{(0-inf)}$ (ng.h/mL) | 710 | 1075 ± 535 | 744 ± 272 |
| AUC$_{cm}$/AUC$_p$ | 0.9 | 1.4 | 1.4 |
| Losartan dosed at 30 mg | | | |
| Losartan | | | |
| $V_{ss}$ (L/kg) | 0.17 | 0.24 | 0.62 ± 0.27 |
| CL (mL/min/kg) | 7.2 | 7.0 ± 1.3 | 8.0 ± 2.7 |
| $t_{1/2}$ (h) | 1.95 | 0.63 ± 0.11 | 3.2 ± 0.8 |
| AUC$_{(0-inf)}$ ng.h/mL | 1197 ± 146 | 1095 ± 212 | 851 ± 253 |
| Carboxylosartan | | | |
| $C_{max}$ (ng/mL) | 108 | 214 ± 95.5 | 149 ± 79 |
| AUC$_{(0-inf)}$ (ng.h/mL) | 1083 | 1588 ± 790 | 1332 ± 490 |
| AUC$_{cm}$/AUC$_p$ | 0.9 | 1.5 | 1.6 |
| Carboxylosartan dosed at 20 mg | | | |
| $V_{ss}$ (L/kg) | 0.16 | 0.21 | 0.14 ± 0.02 |
| CL (mL/min/kg) | 0.36 | 0.72 ± 0.17 | 0.67 ± 0.08 |
| $t_{1/2}$ (h) | 5.0 | 3.73 ± 0.79 | 6.3 ± 1.4 |
| AUC$_{(0-inf)}$ ng.h/mL | 7192 ± 1215 | 7355 ± 1648 | 5969 ± 709 |

[a] Simulated mean results ± SD.
[b] Observed mean data ± SD from study.[9]

b)

| Dose | 0.5 mg/kg BW ( n = 5 ) | | 1 mg/kg BW ( n = 5 ) | |
|---|---|---|---|---|
| Parameters | Plasma | Milk | Plasma | Milk |
| 2 | C max (ng/ml) 15.48 ± 6.64 | 5.34 ± 2.23 | 38.10 ± 8.57 | 11.47 ± 2.23 |
| 3 | T max (d) 0.5 | 0.5 | 0.5 | 0.5 |
| 4 | T 1/2 elim (d) 1.36 ± 0.38 | 1.33 ± 0.37 | 0.95 ± 0.17 | 0.99 ± 0.08 |
| 5 | MRT (d) 1.02 ± 0.12 | 0.97 ± 0.19 | 1.12 ± 0.15 | 1.11 ± 0.14 |
| 6 | AUC (ng d/ml) 17.62 ± 9.68 | 6.56 ± 4.00 | 45.32 ± 13.90 | 13.88 ± 1.77 |
| 7 | AUC milk to plasma ratio 0.36 ± 0.05 | 0.36 ± 0.05 | 0.33 ± 0.08 | 0.33 ± 0.08 |
| 8 | Dose fraction recovered in milk 0.43 ± 0.19 | 0.43 ± 0.19 | 0.42 ± 0.06 | 0.42 ± 0.06 |

c)

| | DoseParameters | 0.5mg/kgBW(n=5)Plasma | 0.5mg/kgBW(n=5)Milk | 1mg/kgBW(n=5)Plasma | 1mg/kgBW(n=5)Milk |
|---|---|---|---|---|---|
| 2 | C max (ng/ml) | 15.48 ± 6.64 | 5.34 ± 2.23 | 38.10 ± 8.57 | 11.47 ± 2.23 |
| 3 | T max (d) | 0.5 | 0.5 | 0.5 | 0.5 |
| 4 | T 1/2 elim (d) | 1.36 ± 0.38 | 1.33 ± 0.37 | 0.95 ± 0.17 | 0.99 ± 0.08 |
| 5 | MRT (d) | 1.02 ± 0.12 | 0.97 ± 0.19 | 1.12 ± 0.15 | 1.11 ± 0.14 |
| 6 | AUC (ng d/ml) | 17.62 ± 9.68 | 6.56 ± 4.00 | 45.32 ± 13.90 | 13.88 ± 1.77 |
| 7 | AUC milk to plasma ratio | 0.36 ± 0.05 | 0.36 ± 0.05 | 0.33 ± 0.08 | 0.33 ± 0.08 |
| 8 | Dose fraction recovered in milk | 0.43 ± 0.19 | 0.43 ± 0.19 | 0.42 ± 0.06 | 0.42 ± 0.06 |

d)

C max ng/ml 15.48 ± 6.64 0.5 mg/kg BW n = 5 _Plasma eprinomectin. ‖ C max ng/ml 5.34 ± 2.23 0.5 mg/kg BW n = 5 _Milk eprinomectin. ‖ C max ng/ml 38.10 ± 8.57 1 mg/kg BW n = 5 _Plasma eprinomectin. ‖ C max ng/ml 11.47 ± 2.23 1 mg/kg BW n = 5 _Milk eprinomectin. ‖ T max d 0.50.5 mg/kg BW n = 5 _Plasma eprinomectin. ‖ T max d 0.50.5 mg/kg BW n = 5 _Milk eprinomectin. ‖ T max d 0.51 mg/kg BW n = 5 _Plasma eprinomectin. ‖ T max d 0.51 mg/kg BW n = 5 _Milk eprinomectin. ‖ T 1/2 elim d 1.36 ± 0.38 0.5 mg/kg BW n = 5 _Plasma eprinomectin. ‖ T 1/2 elim d 1.33 ± 0.37 0.5 mg/kg BW n = 5 _Milk eprinomectin. ‖ T 1/2 elim d 0.95 ± 0.17 1 mg/kg BW n = 5 _Plasma eprinomectin. ‖ T 1/2 elim d 0.99 ± 0.08 1 mg/kg BW n = 5 _Milk eprinomectin. ‖ MRT d1.02 ± 0.12 0.5 mg/kg BW n = 5 _Plasma eprinomectin. ‖ MRT d0.97 ± 0.19 0.5 mg/kg BW n = 5 _Milk eprinomectin. ‖ MRT d1.12 ± 0.15 1 mg/kg BW n = 5 _Plasma eprinomectin. ‖ MRT d1.11 ± 0.14 1 mg/kg BW n = 5 _Milk eprinomectin. ‖ AUC ng d/ml 17.62 ± 9.68 0.5 mg/kg BW n = 5 _Plasma eprinomectin. ‖ AUC ng d/ml 6.56 ± 4.00 0.5 mg/kg BW n = 5 _Milk eprinomectin. ‖ AUC ng d/ml 45.32 ± 13.90 1 mg/kg BW n = 5 _Plasma eprinomectin. ‖ AUC ng d/ml 13.88 ± 1.77 1 mg/kg BW n = 5 _Milk eprinomectin. ‖ AUC milk to plasma ratio 0.36 ± 0.05 0.5 mg/kg BW n = 5 _Plasma eprinomectin. ‖ AUC milk to plasma ratio 0.36 ± 0.05 0.5 mg/kg BW n = 5 _Milk eprinomectin. ‖ AUC milk to plasma ratio 0.33 ± 0.08 1 mg/kg BW n = 5 _Plasma eprinomectin. ‖ AUC milk to plasma ratio 0.33 ± 0.08 1 mg/kg BW n = 5 _Milk eprinomectin. ‖ Dose fraction recovered in milk 0.43 ± 0.19 0.5 mg/kg BW n = 5 _Plasma eprinomectin. ‖ Dose fraction recovered in milk 0.43 ± 0.19 0.5 mg/kg BW n = 5 _Milk eprinomectin. ‖ Dose fraction recovered in milk 0.42 ± 0.06 1 mg/kg BW n = 5 _Plasma eprinomectin. ‖ Dose fraction recovered in milk 0.42 ± 0.06 1 mg/kg BW n = 5 _Milk eprinomectin.

**Figure 4:** Case 3: a) Table extracted from the XML document: https://doi.org/10.1016/j.vetpar.2015.02.013, b) – c) program extracted structured table, d) Sentence format outcome with parameter, value, units – Result of table information extracted as a sentence which has drug name eprinomectin from the table title, appended at the end of each row data. [Note: ‖ indicates separation between table rows].

**Advanced approach:**

A case study considered in the advanced method section is focused on clearance data with drug, dose, animal, route of administration, and clearance as the PK parameters of interest. Here, the table extraction algorithm parses through the XML file types and identifies the table layout. Then, the desired PK parameter, values, units are extracted, save the values in corresponding Python tempfile. Once the desired tables contents are extracted, a .csv file is produced with the given parameters of interest from all the XML tables which can be used for further analysis.

Results for the *doi*s (10.1016/j.ijpharm.2013.12.002 [40], 10.1016/j.xphs.2017.03.032 [38]) include drug, dosage, route of administration, animal, clearance and confirmed the algorithms compatibility to extract the drug and PK parameters of interest from the xml file content. For the table, shown below in *Figure 6a* (Table 2 of doi 10.1016/j.ijpharm.2013.12.002), the outcome includes tempfile to hold the selected parameters, values, followed by its units. It will then be transferred to a .csv file with each row detailing drug, dose, route of administration, animal, clearance. In this case, we have 5 rows for clearance data corresponding to five dosages. Data extracted from the selected table is shown in *Figure 6c*. For Table 3 of article with *doi* 10.1016/j.xphs.2017.03.032 (*Figure 6b*) with different table format is selected as a representative one. Clearance data extracted from the table is shown in *Figure 6d*.

a) Table 2
Pharmacokinetic data of imidol in rats (n = 6).

| Parameters | Dose (mg kg$^{-1}$) | | | | |
|---|---|---|---|---|---|
| (Mean ± SD) | 10.02 (po) | 30.25 (po) | 70.55 (po) | 320.3 (po) | 10.20 (iv) |
| $C_{max}$ (ng ml$^{-1}$) | 64.79 ± 6.44 | 188.1 ± 29.36 | 660.2 ± 174.9 | 4430 ± 1288 | 1896 ± 634 |
| $T_{max}$ (h) | 0.4 ± 0.1 | 0.4 ± 0.1 | 0.4 ± 0.1 | 0.4 ± 0.1 | 0.0 ± 0.0 |
| $AUC_{0-12}$ (ng h$^{-1}$ ml$^{-1}$) | 80.1 ± 7.5 | 269.3 ± 53.4 | 769.4 ± 77.8 | 6130 ± 2063 | 500.8 ± 83.2 |
| $AUC_{0-\infty}$ (ng h$^{-1}$ ml$^{-1}$) | 93.8 ± 11.1 | 287.0 ± 62.2 | 827 ± 97 | 6410 ± 2022 | 534.5 ± 71.76 |
| $CL_z$ (L h$^{-1}$ kg$^{-1}$) | 107.7 ± 11.8 | 108.1 ± .2 | 101.9 ± 24.81 | 540.5 ± 170.7 | 110.6 ± 7.8 |
| $V_z$ (L kg$^{-1}$) | 347.1 ± 229.8 | 525.2 ± 114.7 | 493.5 ± 128.9 | 835 ± 127 | 672.9 ± 289.9 |
| $T_{1/2}$ (h) | 3.4 ± 0.6 | 4.0 ± 1.0 | 4.1 ± 1.4 | 5.4 ± 1.2 | 4.1 ± 1.5 |

b)

Table 3
PK of Carboxylosartan in the Dog Following IV Administration

| ID | Half-life (h) | $V_{ss}$ (L/kg) | $AUC_{0-24hr}$ (ng.h/mL) | $AUC_{0-\infty}$ (ng.h/mL) | $CL_p$ (mL/min/kg) | $f_e$ | $CL_{r,Dog}$ (mL/min/kg) |
|---|---|---|---|---|---|---|---|
| Dog1 | 13 | 1.5 | 1020 | 1040 | 16 | 0.28 | 4.5 |
| Dog2 | 7.4 | 1.1 | 790 | 790 | 21 | 0.25 | 5.3 |
| Mean | 10 | 1.3 | 900 | 920 | 19 | 0.27 | 5.0 |

c)

Drug Imidol, Dose 10.02 mg/kg, route po, animal rat, clearance 107.07±11.8 L/ h/ kg; Drug Imidol, Dose 30.25 mg/kg, route po, animal rat, clearance 108.1±.2 L/ h/ kg; Drug Imidol, Dose 7.55 mg/kg, route po, animal rat, clearance 101.9±24.81 L/ h/ kg; Drug Imidol, Dose 320.3 mg/kg, route po, animal rat, clearance 540.07±170.7 L/ h/ kg; Drug Imidol, Dose 10.02 mg/kg, route iv, animal rat, clearance 110.6±7.8 L/ h/ kg.

d)

Drug Carboxylosartan, Dose 1 mg/kg, route IV, animal Dog1, clearance 16 mL/min/kg; Drug Carboxylosartan, Dose 1 mg/kg, route IV, animal Dog2, clearance 21 mL/min/kg.

**Figure 5:** Table extracted from the XML files for DOIs: a) /10.1016/j.ijpharm.2013.12.002, b) /10.1016/j.xphs.2017.03.032, c) – d) Clearance data outcome with other parameters of interest for the tables a), b) respectively.

## Discussion

Recognizing the structure of a table is important for machines to comprehend and extract relevant data from it. While our algorithm provides a universal process to convert XML tables of varying complexity, there are uncertainties related to the table layout and potential issues with the data itself. The table header is particularly significant in comprehending the table's contents. When it comes to XML table headers, they consist of multiple row tags that can contain header or table data entry tags. The table body also includes both data and header tags. In a way, it increases the complexity of conversion of table objects for further processing in programming languages which may result in inconsistencies in the PK information due to the concurrent use of *thead* tags, or *entry* tags, or a combination of both based on authors choice. Similarly, other table features such as merged-column, merged-row, and spanning cells, also increase the complexity of tables. Even though the proportion of span cells is usually low in a complicated table, they may contain more semantic information when table headers are present in them. Therefore, recognizing complicated table structures is an important part in the text and data mining of tables. The proposed algorithm has the ability to identify intricate table arrangements, recognize table headers through header titles or table data *entry* tags in merged-column, merged-row, or spanning cells, transform them into a data frame, and eventually produce the necessary parameters and values.

In the realm of R&D, literature mining scientific articles is an essential tool for ensuring the reproducibility of research information and facilitating comparison to other studies. While PDF has long been the favored format for electronic scientific articles, XML and HTML file types are growing in popularity due to their flexible information retrieval capabilities. This study aims to develop a system for detecting and extracting tables from XML versions of scientific literature, taking advantage of the structured organization and specific tags assigned to each file element, including images, tables, and text content. The information taken from a table is transformed into a sequence of words so that quantitative data, such as PK parameters, can be extracted and reported. The proposed system involves various steps, including analyzing the table's layout to detect the reading order, identifying, and extracting PK parameters, and storing the data with the appropriate units provided in the table cells. *Table 3* shows some of the table data extraction methods present in literature.

**Table 3:** Literature review of existing table extraction methodologies.

| Data | Source File Type | Description | Reference |
|---|---|---|---|
| Data tables | XML versions of scientific articles | Table data extraction system: table extraction, data formatting, polymer name recognition, property specifier identification, and data extraction | Oka et al. (2021) [12] |
| Image-based tables | PDF articles in PubMed Central Open Access Subset | PubTabNet: image-based table recognition using deep learning models | Zhong et al. (2020) [13] |
| Image-based dataset | Web-based open-domain scientific and business documents | TableBank: automatic localization of table regions based on bounding box | Li et al. (2020) [14] |
| Data tables | PDF and HTML documents | Automatic table data extraction method: data-mining method to automatically extract rapid assay data from structured (table) data | Jaberi et al. (2021) [15] |
| LATEX table snippets | arXiv PDF files | SciTSR: table structure recognition | Chi et al. (2019) [16] |
| Table images | Documents with handwritten and modern table images | From the different table detection and extraction methods discussed, NLPR_PAL showed better performance in both detection and extraction of table data. | Gao et al. (2019) [27] |
| Tables | open-access PDF documents | A benchmarking tool to evaluate the performance of table recognition and extraction methods for neurological disorders research | Adams et al. (2022) [28] |

In addition, we have not only managed to acquire the data from the table, but we have also organized it in a manner that holds significance with further studies. Through our semantic search algorithm, the data is indexed and can be effortlessly located by researchers. This structured pharmacokinetic data facilitates the reproduction and comparison of research with other studies. Furthermore, we can apply this knowledge to predict drug-target interactions, as well as assist in the development of potential drug-repurposing candidates.

Publications often are unsuccessful in offering important details when reporting PK data, which makes it difficult to identify significant PK information and impossible to integrate and reuse the data. For instance, when it comes to drugs, including dosage, strength, or route of administration, in the corresponding table can be beneficial. This way, readers can easily find the information they need without having to search through the entire text. There are additional instances where body weights have not been documented, preventing the ability to convert to dosage per body weight for the purpose of reproducibility. However, we extracted some of that information based on the table title by appending it with table data described in the methodology. After conducting our research, we propose a series of key recommendations for the dissemination of findings from clinical studies results as a table in the field of pharmacokinetics: (i) clearly describe the data and parameters: drug, dose, route of administration [41], and dosage forms [42], the more specific the information, the better it can support future analysis; (ii) publish the actual concentration-time values, in addition to derived or scaled parameters; (iii) provide details about the patient/individual such as basic demographic and anthropometric information including age, species, bodyweight, sex, and height.

Unstructured data is easy to access, but the most important PK data is often presented as tables in scientific publications. As a result, it is essential to develop efficient methods and algorithms for extracting table data from scientific literature. In this study, our aim is to enhance the reporting of pharmacokinetic studies by making data representation and integration more accessible. This will not only improve the reusability of pharmacokinetics information but also simplify the integration of studies and computational models. Therefore, our data extraction methods will enable the availability of data for improved pharmacokinetics research. This study has provided a data-mining method for automatically extracting PK data from XML files which are common in the text domain but with a specific target of table data extraction.

Compared to the rule-based DNN learning polymer data extraction model [12], our approach is significant where the data extraction is not restricted to a limited number of drugs, polymers, or articles. Our drug database contains millions of drugs curated to support advanced pharmacology studies [43]. Other schemes reported in literature focus only on tables depicted as images [13,14,27], which may not have a significant impact on the data collection. Also, when compared to the PDF based table extraction schemes [15,16,28], our approach can easily be modified to handle different file types and table formats.

Nevertheless, data are not always presented in the form of tables. Our current research is also focused on developing a learned information extraction system to convert any text format of PK data into structured data for further processing. PK drugs are used in modern animal agriculture, including recommendations for safe withdrawal intervals of drugs and chemicals in food-producing animals. This data can then be mined for use in animal agriculture. As part of the data mining project, the aim is to create a comprehensive drug database that will assist FARAD in its mission of improving animal health as well as promoting efficient and humane production methods for livestock [43]. The proposed algorithm is an extension of the web crawling system [11] and aims to enhance the FARAD PK database, which was previously a time-consuming and tedious task.

*Supplementary Material S2* displays certain table formats that achieved success and encountered challenges with the proposed algorithm. Proposed table data extraction algorithm does not cover tables depicted as images in scientific articles [44,45]. It is challenging to retrieve the compound or drug names when presented as numbers, for example in the Table 2 of article [46], table title is 'Rat pharmacokinetic data for 8, 10 and 13' and row data is Compound 8, 10, 13, were missing in the data identified and extracted by our table extraction algorithm. Another challenge that we need to tackle is the presence of PK data with both observed and predicted values. For example, in the Table 2 of article [47], the observed and predicted value of 68 compound's PK parameters such as clearance, volume, bioavailability, and AUC, were not reported independently in the curated sentences. Examinations are being conducted to find ways to prevent the loss of significant data from these scenarios.

The present table data extraction study focuses on rule-based approaches for extracting PK parameters. Subsequent work has extended this framework by integrating both large language models (LLMs) and rule-based methods to improve PK parameter extraction without loss of information. These advancements are reported in our later publications [48–51], while this research remains ongoing, with future efforts directed toward comprehensive data extraction across relevant domains.

# Conclusion

This paper examines the automatic machine extraction of PK data from tables in XML versions of scientific articles. In this system, we have learned about and solved important challenges in table detection and extraction from XML file formats focusing on the pharmacokinetic domain. The proposed system is an integration of five steps: table detection, table extraction, reading order detection, PK data recognition with parameter specifier identification, and PK data extraction. Moreover, our understanding of this information has enabled us to enhance the algorithms to obtain a deeper understanding of the inherent reading sequence of the data, where the cells and headers have a predetermined connection. In such situations, comprehending the order of the cell and header features is crucial to transforming tables into meaningful knowledge. In summary, the benefits of our system of automatically extracting and collecting PK parameters from tables include the following: (i) accurately integrating and harmonizing pharmacokinetics information from a wide range of studies and sources; (ii) applicable across a wide range of drugs or compounds and gain insights into how well data is reported in various fields of the table; (iii) decreasing the errors experienced by manual data collection; (iv) improved time and cost effectiveness by limiting end-to-end manual intervention for data collection; (v) ability to handle multiple documents simultaneously and efficiently; (vi) feasibility of real-time implementation of tables extraction methods for dynamic XML sources; and (vii) ability to function for other file formats such as HTML, JSON by identifying and updating specific table tags.

# Acknowledgments

This work was supported by the USDA via the FARAD program and its support for the 1DATA Consortium at Kansas State University. MJ-D also accepted funding from BioNexus KC for this project. Neither USDA nor BioNexus KC had a direct role in this article.

Parts of this manuscript have been previously published in the proceedings of 23rd Biennial Symposium of the American Academy of Veterinary Pharmacology and Therapeutics AAVPT 2025 [48]

[https://cdn.ymaws.com/www.aavpt.org/resource/resmgr/biennial_2025/proceedings_for_the_23rd_aav.pdf, pages 42-47]. This submission represents a significantly revised and expanded version.

**Conflict of Interest**

The authors declare that the research was conducted in the absence of any commercial or financial relationships that could be construed as a potential conflict of interest.

**Table List**

Table 1: Representative PK parameters and their alternatives used in the table data extraction module.

Table 2: a) An example of a complex table with spanning cells in column and row data fields, and b) complete table structure generated by using the proposed algorithm.

Table 3: Literature review of existing table extraction methodologies.

**Figure List**

Figure 1: Overall workflow of table data extraction methodology from scientific articles of XML and HTML file types.

Figure 2: Common table formats. These tables were obtained from the following XML documents: a) 10.1016/j.smallrumres.2018.01.001, b) 10.1016/j.xphs.2017.03.032, c) 10.1016/j.vetpar.2015.02.013, d) 10.1016/j.xphs.2017.03.032.

Figure 3: Flow chart representing the systematic workflow of table extraction algorithms from scraping the tables to extracting results in a csv file format for research purposes.

Figure 4: Case 1: a) Table extracted from the following XML document: https://doi.org/10.1016/j.smallrumres.2018.01.001, b) program extracted table, c) Result of table information extracted as a sentence which has drug name MEL from the table title, appended at the end of each row data. [Note: ‖ indicates separation between table rows] Sentence format outcome with parameter, value, units.

Figure 5: Case 3: a) Table extracted from the XML document: https://doi.org/10.1016/j.vetpar.2015.02.013, b) – c) program extracted structured table, d) Sentence format outcome with parameter, value, units – Result of table information extracted as a sentence which has drug name eprinomectin from the table title, appended at the end of each row data. [Note: ‖ indicates separation between table rows].

Figure 6: Table extracted from the XML files for DOIs: a) /10.1016/j.ijpharm.2013.12.002, b) /10.1016/j.xphs.2017.03.032, c) – d) Clearance data outcome with other parameters of interest for the tables a) – b) respectively.

**Automated Extraction of Pharmacokinetic Parameters from Structured XML Scientific Articles: Enhancing Data Accessibility at Scale**

Remya Ampadi Ramachandran[1,2,3], Lisa A. Tell[4], Sid Sidharth[1,2,3], Nuwan Millagaha Gedara[1], Hossein Sholehrasa[1,2,5], Jim E. Riviere[1,2], Majid Jaberi-Douraki[1,2,3,*]

[1]1DATA Consortium, www.1DATA.life, Kansas State University Olathe, Olathe, KS, USA.

[2]Food Animal Residue Avoidance and Databank Program (FARAD), Kansas State University Olathe, Olathe, KS, USA.

[3]Department of Mathematics, Kansas State University, Manhattan, KS, United States

[4]FARAD, Department of Medicine and Epidemiology, School of Veterinary Medicine, University of California-Davis, Davis, CA

[5]Department of Computer Science, Kansas State University, Manhattan, KS, United States

[*] Corresponding Author: jaberi@k-state.edu

**Supplementary Material S2**

**Section A:** In this section we have included certain table formats that were successful in terms of PK data extraction with the implemented algorithm [1–13].

A) 10.1016/j.smallrumres.2018.01.001

**Table 1**
Pharmacokinetic parameters of MEL after IV administration at 0.5 mg/kg in lactating goats (n = 6).

| Parameter | Units | IV Mean SD |
|---|---|---|
| AUC | h*ng/mL | 26499 ± 4233 |
| K10 | 1/h | 0.12 ± 0.03 |
| K12 | 1/h | 0.64 ± 0.38 |
| K21 | 1/h | 1.13 ± 0.71 |
| K10_HL | h | 6.07 ± 1.18 |
| Alpha | 1/h | 1.82 ± 1.09 |
| Beta | 1/h | 0.07 ± 0.02 |
| Alpha_HL | h | 0.53 ± 0.35 |
| Beta_HL | h | 9.96 ± 2.51 |
| A | ng/mL | 1223 ± 153.71 |
| B | ng/mL | 1840 ± 357.69 |
| AUMC | h*h*ng/mL | 374373 ± 120223 |
| MRT | h | 13.88 ± 3.36 |
| CL | mL/h/kg | 19.38 ± 3.86 |
| Vss | mL/kg | 262.37 ± 50.74 |
| V1 | mL/kg | 165.76 ± 23.06 |
| V2 | mL/kg | 96.61 ± 31.07 |

Area under the curve (AUC), elimination rate from compartment 1 (K10), rate of movement from compartment 1–2 (K12), the rate of movement from compartment 2–1 (K21), half-life of the elimination phase (K10_HL), rate constant associated with distribution ($\alpha$), rate constant associated with elimination ($\beta$), distribution half-life (Alpha _HL), elimination half-life (Beta_HL), intercept for the distribution phase (A), intercept for the elimination phase (B), area under the first moment curve (AUMC), mean resident time (MRT); total clearance (CL), volume of distribution at the steady state (Vss), volume of compartment 1 (V1), and volume of compartment 2 (V2).

B) 10.1016/j.vetpar.2015.02.013

**Table 1**
Selected pharmacokinetic parameters describing the disposition of topical eprinomectin in plasma and milk after oral administration (0.5 and 1 mg/kg BW) to lactating goats. Values are the mean ± standard deviation [$C_{max}$: observed peak plasma concentration; $t_{max}$: time to reach $C_{max}$; $t_{1/2\ elim}$: half life of elimination; MRT: mean residence time; AUC: area under the plasma or milk concentration vs time curve].

| Dose | 0.5 mg/kg BW (n=5) | | 1 mg/kg BW (n=5) | |
|---|---|---|---|---|
| Parameters | Plasma | Milk | Plasma | Milk |
| $C_{max}$ (ng/ml) | 15.48±6.64[a] | 5.34±2.23[A] | 38.10±8.57[b] | 11.47±2.23[B] |
| $T_{max}$ (d) | 0.5 | 0.5 | 0.5 | 0.5 |
| $T_{1/2\ elim}$ (d) | 1.36±0.38[a] | 1.33±0.37[A] | 0.95±0.17[a] | 0.99±0.08[A] |
| MRT (d) | 1.02±0.12[a] | 0.97±0.19[A] | 1.12±0.15[a] | 1.11±0.14[A] |
| AUC (ng d/ml) | 17.62±9.68[a] | 6.56±4.00[A] | 45.32±13.90[b] | 13.88±1.77[B] |
| AUC milk to plasma ratio | | 0.36±0.05[a] | | 0.33±0.08[a] |
| Dose fraction recovered in milk | | 0.43±0.19[a] | | 0.42±0.06[a] |

Means in the same row with different small-letter (a, b) or capital-letter (A, B) superscripts are significantly different ($P<0.05$).

## C) 10.1016/j.ijpharm.2013.12.002

Table 2
Pharmacokinetic data of imidol in rats (n = 6).

| Parameters (Mean ± SD) | Dose (mg kg$^{-1}$) | | | | |
|---|---|---|---|---|---|
| | 10.02 (po) | 30.25 (po) | 70.55 (po) | 320.3 (po) | 10.20 (iv) |
| $C_{max}$ (ng ml$^{-1}$) | 64.79 ± 6.44 | 188.1 ± 29.36 | 660.2 ± 174.9 | 4430 ± 1288 | 1896 ± 634 |
| $T_{max}$ (h) | 0.4 ± 0.1 | 0.4 ± 0.1 | 0.4 ± 0.1 | 0.4 ± 0.1 | 0.0 ± 0.0 |
| $AUC_{0-12}$ (ng h$^{-1}$ ml$^{-1}$) | 80.1 ± 7.5 | 269.3 ± 53.4 | 769.4 ± 77.8 | 6130 ± 2063 | 500.8 ± 83.2 |
| $AUC_{0-\infty}$ (ng h$^{-1}$ ml$^{-1}$) | 93.8 ± 11.1 | 287.0 ± 62.2 | 827 ± 97 | 6410 ± 2022 | 534.5 ± 71.76 |
| $CL_z$ (L h$^{-1}$ kg$^{-1}$) | 107.7 ± 11.8 | 108.1 ± .2 | 101.9 ± 24.81 | 540.5 ± 170.7 | 110.6 ± 7.8 |
| $V_z$ (L kg$^{-1}$) | 347.1 ± 229.8 | 525.2 ± 114.7 | 493.5 ± 128.9 | 835 ± 127 | 672.9 ± 289.9 |
| $T_{1/2}$ (h) | 3.4 ± 0.6 | 4.0 ± 1.0 | 4.1 ± 1.4 | 5.4 ± 1.2 | 4.1 ± 1.5 |

## D) 10.1016/j.xphs.2017.03.032

Table 3
PK of Carboxylosartan in the Dog Following IV Administration

| ID | Half-life (h) | $V_{ss}$ (L/kg) | $AUC_{0-24hr}$ (ng.h/mL) | $AUC_{0-\infty}$ (ng.h/mL) | $CL_p$ (mL/min/kg) | $f_e$ | $CL_{r,Dog}$ (mL/min/kg) |
|---|---|---|---|---|---|---|---|
| Dog1 | 13 | 1.5 | 1020 | 1040 | 16 | 0.28 | 4.5 |
| Dog2 | 7.4 | 1.1 | 790 | 790 | 21 | 0.25 | 5.3 |
| Mean | 10 | 1.3 | 900 | 920 | 19 | 0.27 | 5.0 |

## E) 10.1016/j.ejca.2008.10.022

Table 4 – Pharmacokinetic parameters.

| | Dose level 1 (30 mg/d) Cisplatin (n = 4) Gemcitabine (n = 2) | | | | Dose level 2 (45 mg/d) Cisplatin (n = 3) Gemcitabine (n = 6) | | | |
|---|---|---|---|---|---|---|---|---|
| | Clearance (l/h) | | Vd (l) | | Clearance (l/h) | | Vd (l) | |
| | Cycle 1 | Cycle 2 | Cycle 1 | Cycle 2 | Cycle 1 | Cycle 2 | Cycle 1 | Cycle 2 |
| Cisplatin | 21.9 ± 2.5 | 21.1 ± 6.1 | 22.8 ± 11.9 | 20.6 ± 9.6 | 17.2 ± 1.7 | 19.0 ± 3.6 | 20.9 ± 5.6 | 18.1 ± 3.8 |
| Gemcitabine | 124 ± 14.4 | 93.4[a] ± 8.0 | 148 ± 144 | 204 ± 126 | 123 ± 25.1 | 114[a] ± 37.1 | 151 ± 80.0 | 134 ± 83.2 |
| Cediranib | $C_{ss\ max}$ = 66.9 ng/ml $AUC_{ss}$ = 1058.7 ng * h/ml | | | | $C_{ss\ max}$ = 194 ng/ml $AUC_{ss}$ = 2861.7 ng * h/ml | | | |

Key: l, litres; Vd, volume of distribution; h, hour; AUC, area under the concentration–time curve; ss, steady state.
a  p < 0.02 versus cycle 1.

## F) 10.1016/S0378-5173(01)00654-8

Table 1
Main pharmacokinetic parameters following SuBP and pamidronate administration (mean ± S.D.)

| Drug | Administration | AUC 24h (mcg min/ml) | CL (ml/min per kg) | $CL_r$ (ml/min per kg) | $CL_{TIBIA}$ (ml/min per kg) | $T_{1/2}$ (min) | $V_{SS}$ (ml/kg) |
|---|---|---|---|---|---|---|---|
| SuBP | IV bolus 1 mg/kg (iso-osmotic) | 97.8 ± 21.7 | 10.6 ± 2.3 | 5.97 ± 2.59 | 0.040 ± 0.008 | 21.7 ± 4.9 | 307 ± 117 |
| SuBP | IV bolus 1 mg/kg (hypo-osmotic) | 160 ± 13 | 5.88 ± 5.23 | 1.44 ± 0.95 | 0.045 ± 0.004 | 173 ± 51 | 864 ± 209 |
| SuBP | IV infusion 4h 1 mg/kg | 144 ± 21 | 7.12 ± 1.12 | 5.74 ± 0.63 | 0.028 ± 0.005 | – | 202 ± 32 |
| SuBP | PO 10 mg/kg | 428 ± 92 | – | 1.02 ± 0.59 | 0.0028 ± 0.0008 | 288 ± 104 | – |
| SuBP | PO 40 mg/kg | 1164 ± 414 | – | 0.50 ± 0.21 | 0.0056 ± 0.0023 | 308 ± 156 | – |
| Kruskal–Wallis test (P) | | – | <0.01 | <0.01 | <0.01 | <0.01 | <0.05 |
| Pamidronate | IV bolus 1 mg/kg (iso-osmotic) | 66.3 ± 7.9 | 15.2 ± 3.5 | 1.92 ± 0.75 | 0.27 ± 0.03 | 17.2 ± 6.6 | 183 ± 48 |
| Pamidronate | IV bolus 1 mg/kg (hypo-osmotic) | 62.9 ± 12.6 | 16.4 ± 3.1 | 1.24 ± 0.41 | 0.28 ± 0.07 | 22.2 ± 14.7 | 273 ± 45 |
| Pamidronate | IV infusion 4h 1 mg/kg | 185 ± 19 | 5.45 ± 0.53 | 0.37 ± 0.18 | 0.081 ± 0.022 | – | 64.8 ± 6.6 |
| Pamidronate | PO 10 mg/kg | 43.0 ± 21.2 | – | 0.43 ± 0.21 | 0.011 ± 0.006 | 105 ± 38 | – |
| Kruskal–Wallis test (P) | | – | <0.01 | <0.05 | <0.01 | <0.01 | <0.01 |

G) 10.1016/S0928-0987(96)00254-0

Table 2
Summary of the NONMEM analyses carried out with a one compartment multiple dose model with first order absorption

| Model | Parameters | Parameter estimates[a] | Interindividual variability[b] | Objective function | Intraindividual variability[d] (mg/l) |
|---|---|---|---|---|---|
| Model I Base[c] | CL/F (l/h) V/F (l) | 2.26 (0.11) 97.1 (32.0) | 27.4% 80.1% | 195.757 | 0.660 |
| Model II Base with CL and V∝to body weight | CL/F (l/h/kg) V/F (l/kg) | 0.0321 (0.0013) 1.40 (0.20) | 24.2% 80.4% | 183.152 | 0.664 |
| Model III[e] Model II with CL and V split on sex (M — male/F — female) | CL/F (l/h/kg — M) CL/F (l/h/kg — F) V/F (l/kg — M) V/F (l/kg — F) | 0.0377 0.0289 1.67 1.05 | 20.2% 74.7% | 172.716 | 0.670 |

[a]Values in parentheses are standard errors.
[b]Interindividual variability expressed as a percentage.
[c]Base model refers to one compartment model with no covariates.
[d]Intraindividual variability is the square root of the residual variability described by an additive term in the model.
[e]NONMEM failed to converge successfully and no estimates of standard error were produced for this model. The same intra- and interindividual variance term was used for both sexes.

H) 10.1016/bs.pmch.2018.01.001

Table 1 Rat iv Pharmacokinetic Parameters of Compounds **8**, **9**, **10** and Ivacaftor (**1**)

| Compound | Clearance (Clp, mL/min/kg) | Half-Life ($t_{1/2}$, h) | Volume of Distribution ($V_{ss}$, L/kg) |
|---|---|---|---|
| **8** | 85.6 | 0.7 | 2.9 |
| **9** | 62.9 | 1.1 | 2.9 |
| **10** | 17.7 | 2.9 | 3.2 |
| **1** (Ivacaftor) | 5.5 | 9.5 | 3.6 |

I) doi:10.1002/jps.10427

Table 3. Pharmacokinetic Parameters of KR-60436 at 20 mg/kg in Rats after 30-Minute Intravenous (iv) and Intraportal (ip) Administrations and after Intraportal (ip), Intragastric (ig), and Intraduodenal (id) Administration[a]

| Parameter | iv (n = 5) | ip (n = 5) | ip (n = 3) | ig (n = 3) | id (n = 4) |
|---|---|---|---|---|---|
| Body weight (g) | 278 ± 24.1 | 278 ± 28.8 | 308 ± 17.5 | 290 ± 7.07 | 311 ± 13.8 |
| AUC (μg · min/mL) | 268 ± 55.3[b] | 208 ± 21.7 | 322 ± 30.3[c] | 56.7 ± 8.00 | 137 ± 74.0 |
| Terminal half-life (min) | 176 ± 61.9[b] | 126 ± 29.7 | 223 ± 21.3 | 195 ± 95.4 | 200 ± 15.0 |
| CL (mL/min/kg) | 74.5 ± 22.4[b] | 96.1 ± 15.4 | — | — | — |

[a]Values expressed as mean ± SD.
[b]Intravenous (iv) was significantly different ($p < 0.05$) from ip.
[c]Each route of administration was significantly different ($p < 0.05$).

## J) 10.1016/j.ijantimicag.2011.07.013

| CVVHF | Dose (mg/kg) | AUC (mgh/L) | CL$_{total}$ (L/h/kg) | V$_{d,ss}$ (L/kg) | T$_{1/2}$ (h) | Dilution mode | Membrane/surface area (m$^2$) | Q$_{bf}$ (mL/min) | Q$_{uf}$ (ml |
|---|---|---|---|---|---|---|---|---|---|
| Bellmann et al. [54] | | 3.25 | 9.59 | 0.274 | 5.89 | | 21.04 | | |
| Bellmann et al. [54] | 2.82 | 5.3 | 0.389 | 9.17 | 25.29 | Pre | 0.71 Polysulfone | 168 | 204 |

## K) 10.1016/S0306-3623(98)00013-5

TABLE 1. Effect of water deprivation in Nubian goats ($n = 6$) on the pharmacokinetic parameters of oxytetracycline hydrochloride after intravenous injection at a dose rate of 5 mg/kg body weight

| Pharmacokinetic parameter | Normal | Dehydrated (7.6% weight loss) | (10.3% weight loss) | (12.7% weight loss) |
|---|---|---|---|---|
| Cop (µg/ml)[1] | 4.74 ± 0.57 | 4.12 ± 0.43 | 5.15 ± 0.56 | 6.50 ± 0.91 |
| A (µg/ml) | 3.62 ± 0.43 | 2.50 ± 0.26 | 3.35 ± 0.33 | 5.51 ± 0.78 |
| B (µg/ml) | 1.11 ± 0.15 | 1.63 ± 0.17* | 1.80 ± 0.23* | 0.98 ± 0.14 |
| $t_{1/2\alpha}$ (hr) (harmonic mean) | 0.35 | 0.73 | 2.10 | 2.23 |
| $t_{1/2\beta}$ (hr) (harmonic mean) | 3.99 | 5.53 | 6.59 | 9.19 |
| α (1/hr) | 2.005 ± 0.197 | 0.955 ± 0.144** | 0.329 ± 0.042*** | 0.312 ± 0.040*** |
| β (1/hr) | 0.174 ± 0.023 | 0.125 ± 0.018 | 0.105 ± 0.015* | 0.075 ± 0.010** |
| $k_{12}$ (1/hr) | 0.999 ± 0.086 | 0.363 ± 0.057*** | 0.063 ± 0.007*** | 0.064 ± 0.008*** |
| $k_{21}$ (1/hr) | 0.605 ± 0.069 | 0.453 ± 0.067 | 0.184 ± 0.026*** | 0.111 ± 0.034*** |
| $k_{el}$ (1/hr) | 0.575 ± 0.068 | 0.264 ± 0.039** | 0.189 ± 0.024*** | 0.212 ± 0.027*** |
| AUC (µg/ml/hr) | 8.23 ± 0.08 | 16.04 ± 0.71*** | 27.61 ± 0.64*** | 30.54 ± 0.46*** |
| $V_c$ (ml/kg) | 1141.02 ± 162.70 | 1280.77 ± 149.99 | 1033.93 ± 131.54 | 849.67 ± 132.47 |
| Cl$_{total}$ (ml/kg/hr) | 608.18 ± 5.96 | 314.61 ± 14.42*** | 181.56 ± 4.47*** | 163.88 ± 2.38*** |
| $V_{dss}$ (ml/kg) | 3088.09 ± 491.50 | 2294.98 ± 258.70 | 1399.98 ± 188.11** | 1341.97 ± 210.66*** |

## L) 10.1016/j.actatropica.2008.05.013

Table 1
Piperaquine pharmacokinetic parameters for fed and fasting healthy subjects after an oral single dose of piperaquine phosphate and dihydroartemisinin determined by noncompartmental analysis

| Parameter | Fed ($n = 16$) | | Fasting ($n = 16$) | | Mann–Whitney |
|---|---|---|---|---|---|
| | Median | 80% central range | Median | 80% central range | P |
| C$_{max}$ (µg/L) | 212 | 130–368 | 130 | 50–407 | 0.22 |
| T$_{max}$ (h) | 4 | 3–9 | 4 | 1–9 | 0.54 |
| AUC$_{0-24}$ (h mg/L) | 2.2 | 1.4–3.6 | 1.7 | 0.7–3.6 | 0.13 |
| AUC$_{0-last}$ (h mg/L) | 11.5 | 6.9–17.3 | 13.9 | 2.8–19.3 | 0.36 |
| AUC$_{0-\infty}$ (h mg/L) | 20.9 | 8.8–40.5 | 23.1 | 3.9–49.5 | 0.78 |
| CL/F (L/(h kg)) | 0.3 | 0.13–0.69 | 0.27 | 0.13–1.61 | 0.87 |
| V$_z$/F (L/kg) | 193 | 146–390 | 262 | 93–483 | 0.29 |
| $t_{1/2\lambda_z}$ (day) | 18 | 6–72 | 20 | 5–101 | 0.98 |

M) 10.1016/j.jchromb.2021.122862

Table 3. Pharmacokinetic (PK) parameters of niclosamide after intravenous, oral, and intramuscular administration to rats and dogs.

| PK Parameter | Sprague Dawley Rat | | | | | Beagle Dog |
|---|---|---|---|---|---|---|
| | Intravenous | | | Oral | Intramuscular | Intravenous |
| Dose (mg/kg) | 0.3 | 1 | 3 | 1 | 1 | 2 |
| $T_{max}$ (h) | 0.083±0.000 | 0.083±0.000 | 0.083±0.000 | 0.950±0.671 | 0.083±0.000 | 0.083±0.000 |
| $C_{max}$ (ng/mL) | 1035±166.4 | 3088±480.5 | 11920±1144 | 22.42±7.992 | 1566±97.88 | 2543±385.5 |
| $T_{1/2}$ (h) | 1.006±0.143 | 1.247±0.445 | 1.056±0.542 | 1.820±0.581 | 1.174±0.231 | 1.030±0.893 |
| $AUC_{last}$ (ng·h/mL) | 300.9±49.17 | 902.5±150.3 | 3375±254.3 | 44.71±11.05 | 585.7±86.65 | 811.5±120.8 |
| $AUC_{inf}$ (ng·h/mL) | 302.6±48.28 | 905.0±150.1 | 3378±254.0 | 49.88±9.242 | 589.2±86.29 | 813.1±122.0 |
| CL (mL/(h·kg)) | 1012±163.8 | 1130±187.7 | 892.2±66.86 | NC | NC | 2496±360 |
| $V_{ss}$ (mL/kg) | 303.8±133.2 | 256.2±59.80 | 170.7±77.07 | NC | NC | 661.2±61.7 |
| Bioavailability (%) | | | | 5.512±1.021 | 65.10±9.535 | 100 |

**Section B:** In this section we have included certain table formats that were successful but with some challenges in terms of the extracted PK data with the implemented algorithm. In future studies, we aim to tackle challenges that come under these scenarios [2,7,9,11].

A) 10.1016/j.ejca.2008.10.022

**Table 4 – Pharmacokinetic parameters.**

| | Dose level 1 (30 mg/d) Cisplatin (n = 4) Gemcitabine (n = 2) | | | | Dose level 2 (45 mg/d) Cisplatin (n = 3) Gemcitabine (n = 6) | | | |
|---|---|---|---|---|---|---|---|---|
| | Clearance (l/h) | | Vd (l) | | Clearance (l/h) | | Vd (l) | |
| | Cycle 1 | Cycle 2 | Cycle 1 | Cycle 2 | Cycle 1 | Cycle 2 | Cycle 1 | Cycle 2 |
| Cisplatin | 21.9 ± 2.5 | 21.1 ± 6.1 | 22.8 ± 11.9 | 20.6 ± 9.6 | 17.2 ± 1.7 | 19.0 ± 3.6 | 20.9 ± 5.6 | 18.1 ± 3.8 |
| Gemcitabine | 124 ± 14.4 | 93.4$^a$ ± 8.0 | 148 ± 144 | 204 ± 126 | 123 ± 25.1 | 114$^a$ ± 37.1 | 151 ± 80.0 | 134 ± 83.2 |
| Cediranib | $C_{ss\ max}$ = 66.9 ng/ml $AUC_{ss}$ = 1058.7 ng*h/ml | | | | $C_{ss\ max}$ = 194 ng/ml $AUC_{ss}$ = 2861.7 ng*h/ml | | | |

Key: l, litres; Vd, volume of distribution; h, hour; AUC, area under the concentration–time curve; ss, steady state.
a  p < 0.02 versus cycle 1.

> In this table, $AUC_{ss}$ values such as 1058.7 ng * h/ml, and 2861.7 ng * h/ml for the drug Cediranib is considered as clearance values given the header row is Clearance

B) 10.1016/S0378-5173(01)00654-8

Table 1
Main pharmacokinetic parameters following SuBP and pamidronate administration (mean ± S.D.)

| Drug | Administration | AUC 24h (mcg min/ml) | CL (ml/min per kg) | CL$_r$ (ml/min per kg) | CL$_{TIBIA}$ (ml/min per kg) | $T_{1/2}$ (min) | $V_{SS}$ (ml/kg) |
|---|---|---|---|---|---|---|---|
| SuBP | IV bolus 1 mg/kg (iso-osmotic) | 97.8 ± 21.7 | 10.6 ± 2.3 | 5.97 ± 2.59 | 0.040 ± 0.008 | 21.7 ± 4.9 | 307 ± 117 |
| SuBP | IV bolus 1 mg/kg (hypo-osmotic) | 160 ± 13 | 5.88 ± 5.23 | 1.44 ± 0.95 | 0.045 ± 0.004 | 173 ± 51 | 864 ± 209 |
| SuBP | IV infusion 4h 1 mg/kg | 144 ± 21 | 7.12 ± 1.12 | 5.74 ± 0.63 | 0.028 ± 0.005 | – | 202 ± 32 |
| SuBP | PO 10 mg/kg | 428 ± 92 | – | 1.02 ± 0.59 | 0.0028 ± 0.0008 | 288 ± 104 | – |
| SuBP | PO 40 mg/kg | 1164 ± 414 | – | 0.50 ± 0.21 | 0.0056 ± 0.0023 | 308 ± 156 | – |
| Kruskal–Wallis test (P) | | – | <0.01 | <0.01 | <0.01 | <0.01 | <0.05 |
| Pamidronate | IV bolus 1 mg/kg (iso-osmotic) | 66.3 ± 7.9 | 15.2 ± 3.5 | 1.92 ± 0.75 | 0.27 ± 0.03 | 17.2 ± 6.6 | 183 ± 48 |
| Pamidronate | IV bolus 1 mg/kg (hypo-osmotic) | 62.9 ± 12.6 | 16.4 ± 3.1 | 1.24 ± 0.41 | 0.28 ± 0.07 | 22.2 ± 14.7 | 273 ± 45 |
| Pamidronate | IV infusion 4h 1 mg/kg | 185 ± 19 | 5.45 ± 0.53 | 0.37 ± 0.18 | 0.081 ± 0.022 | – | 64.8 ± 6.6 |
| Pamidronate | PO 10 mg/kg | 43.0 ± 21.2 | – | 0.43 ± 0.21 | 0.011 ± 0.006 | 105 ± 38 | – |
| Kruskal–Wallis test (P) | | – | <0.01 | <0.05 | <0.01 | <0.01 | <0.01 |

> Some drug names shown as abbreviations in the table, like **_SuBP_** in this case was missing in the curated dataset.

C) 10.1016/S0928-0987(96)00254-0

Table 2
Summary of the NONMEM analyses carried out with a one compartment multiple dose model with first order absorption

| Model | Parameters | Parameter estimates[a] | Interindividual variability[b] | Objective function | Intraindividual variability[d] (mg/l) |
|---|---|---|---|---|---|
| Model I Base[c] | CL/F (l/h) V/F (l) | 2.26 (0.11) 97.1 (32.0) | 27.4% 80.1% | 195.757 | 0.660 |
| Model II Base with CL and V∝to body weight | CL/F (l/h/kg) V/F (l/kg) | 0.0321 (0.0013) 1.40 (0.20) | 24.2% 80.4% | 183.152 | 0.664 |
| Model III[e] Model II with CL and V split on sex (M — male/F — female) | CL/F (l/h/kg — M) CL/F (l/h/kg — F) V/F (l/kg — M) V/F (l/kg — F) | 0.0377 0.0289 1.67 1.05 | 20.2% 74.7% | 172.716 | 0.670 |

[a] Values in parentheses are standard errors.
[b] Interindividual variability expressed as a percentage.
[c] Base model refers to one compartment model with no covariates.
[d] Intraindividual variability is the square root of the residual variability described by an additive term in the model.
[e] NONMEM failed to converge successfully and no estimates of standard error were produced for this model. The same intra- and interindividual variance term was used for both sexes.

➢ In this table, by considering the shaded regions as merged cells, our algorithm filled in the values based on the previous cell.

D) 10.1016/bs.pmch.2018.01.001

Table 1 Rat iv Pharmacokinetic Parameters of Compounds 8, 9, 10 and Ivacaftor (1)

| Compound | Clearance (Clp, mL/min/kg) | Half-Life ($t_{1/2}$, h) | Volume of Distribution ($V_{ss}$, L/kg) |
|---|---|---|---|
| 8 | 85.6 | 0.7 | 2.9 |
| 9 | 62.9 | 1.1 | 2.9 |
| 10 | 17.7 | 2.9 | 3.2 |
| 1 (Ivacaftor) | 5.5 | 9.5 | 3.6 |

➢ Our algorithm is limited to capture compounds when expressed as numbers.